# SPATIOTEMPORAL ADAPTIVE QUANTIZATION FOR THE PERCEPTUAL VIDEO CODING OF RGB 4:4:4 DATA

*Lee Prangnell and Victor Sanchez*
{lee.prangnell, v.f.sanchez-silva}@warwick.ac.uk

Department of Computer Science, University of Warwick, England, UK

## ABSTRACT

Due to the spectral sensitivity phenomenon of the Human Visual System (HVS), the color channels of raw RGB 4:4:4 sequences contain significant psychovisual redundancies; these redundancies can be perceptually quantized. The default quantization systems in the HEVC standard are known as Uniform Reconstruction Quantization (URQ) and Rate Distortion Optimized Quantization (RDOQ); URQ and RDOQ are not perceptually optimized for the coding of RGB 4:4:4 video data. In this paper, we propose a novel spatiotemporal perceptual quantization technique named SPAQ. With application for RGB 4:4:4 video data, SPAQ exploits HVS spectral sensitivity-related color masking in addition to spatial masking and temporal masking; SPAQ operates at the Coding Block (CB) level and the Prediction Unit (PU) level. The proposed technique perceptually adjusts the Quantization Step Size (QStep) at the CB level if high variance spatial data in G, B and R CBs is detected and also if high motion vector magnitudes in PUs are detected. Compared with anchor 1 (HEVC HM 16.17 RExt), SPAQ considerably reduces bitrates with a maximum reduction of approximately 80%. The Mean Opinion Score (MOS) in the subjective evaluations, in addition to the SSIM scores, show that SPAQ successfully achieves perceptually lossless compression compared with anchors.

## 1. INTRODUCTION

Due to an increasing consumer demand for a high fidelity visual experience, the utilization of RGB 4:4:4 video data is becoming ubiquitous in various applications including digital cinema, computer vision, machine learning, medical telepathology, home entertainment and video conferencing. Therefore, to attain a high brightness, hue and saturation fidelity experience for the human observer, the direct coding of RGB 4:4:4 data is becoming more prevalent on, for example, High Definition and Ultra HD displays that are capable of the technical capabilities specified in ITU-R Recommendation BT.2020 [1] and BT.2100 [2]. This includes High Dynamic Range (HDR) tone mapping, Wide Color Gamut (WCG) and high bit-depth (deep color) RGB 4:4:4 playback. Support for the direct coding of RGB 4:4:4 data is included in the HEVC standard [3, 4] including JCT-VC standardized Range Extensions of HEVC HM (HM RExt) [5] and the Screen Content Coding Extensions of HM RExt (HM RExt + SCM) [6]. This includes the coding of RGB 4:4:4 data of up to 16-bits per sample (i.e., deep color RGB) [1, 2]. Due to the aforementioned HVS spectral sensitivity to photons that are perceived as green, the HEVC standard, by default, treats RGB data in the order of G, B and R. That is, the G channel is treated as the most important perceptual channel; this is similar to the way in which Y is treated as the most important perceptual channel in YCbCr data.

As regards the coding of RGB 4:4:4 video data in HEVC, the raw data is partitioned into Coding Units (CUs), which consist of three equal sized CBs (i.e., a Red CB, a Green CB and a Blue CB) within each CU [7]. Coding techniques, including spatiotemporal prediction [8], transform coding [9], quantization [4, 10, 11] and lossless entropy coding [12], operate in the same manner for both RGB 4:4:4 video data and YCbCr 4:4:4 video data. The main scalar quantization techniques in HEVC, known as URQ [4, 10] and RDOQ [11], are both always enabled by default; however, they are not perceptually optimized. URQ is designed to indiscriminately quantize transform coefficients in G, B and R (or Y, Cb and Cr) Transform Blocks (TBs) at equal levels according to the QStep; the QStep is controlled by a Quantization Parameter (QP) [4, 10]. RDOQ, which is utilized in combination with URQ, is a coefficient-level method designed to quantify quantization induced distortion and the number of bits required to encode a quantized coefficient. RDOQ chooses an optimal coefficient value, which is subsequently determined by ascertaining an appropriate trade-off between the bitrate and the distortion; this is known as the rate-distortion cost [11].

In the past decade, several video compression algorithms have been proposed for the direct coding of RGB 4:4:4 data. These techniques include those proposed by Song et al. [13], Kim et al. [14], Zhao and Ai [15], Huang et al. [16], Huang and Lei [17] and Shang et al. [18]. The methods proposed in [13]-[17] are designed primarily to reduce statistical redundancies only. In other words, psychovisual redundancies in raw RGB 4:4:4 data are not taken into account; therefore, perceptual optimization is neglected thus potentially leading to a waste of bits during the residual coding process. The most recent relevant technique proposed (Shang et al. [18] in 2019) is a coefficient-level perceptual quantization method for HEVC; this method employs Quantization Matrices (QMs). The main drawbacks of this method are as follows: firstly, it is not perceptually adaptive (i.e., the QMs are static and do not adapt to the raw data); secondly, it focuses on spatial masking only.

To reiterate, in this paper we propose a novel spatiotemporal perceptually adaptive quantization method, named SPAQ. SPAQ exploits HVS psychovisual redundancies inherent in raw RGB 4:4:4 video data. More specifically we employ color masking by accounting for spectral sensitivity of the HVS (i.e., quantizing data in the B and R channels more coarsley), spatial masking in high variance regions of the raw data and PU-level temporal masking in high motion regions of the raw data. Our technique adaptively discards perceptual redundancies in each color channel from the raw RGB 4:4:4 data by virtue of CB-level perceptual quantization. Compared with the aforementioned previously proposed methods, our technique provides some key advantages, which are as follows: 1) SPAQ employs color masking, spatial masking and temporal masking, and 2) SPAQ is adaptive; therefore, every sequence is perceptually compressed according to the unique characteristics of the sequence being processed.

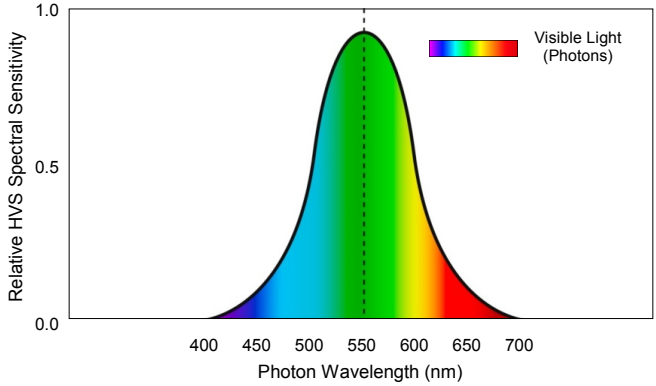

**Fig. 1**: The relative spectral sensitivity of the HVS to photons of various wavelengths (data: National Physical Laboratory [19]). Note how, in terms of the perceived brightness, saturation and hue of color, the HVS is significantly more sensitive to the photon energies that the HVS interprets as the color green.

The rest of this paper is organized as follows. Section 2 includes a detailed overview of the relevant literature. Section 3 provides detailed background information on the SPAQ algorithm. Section 4 includes the evaluation, results and discussion of the proposed method. Finally, section 5 concludes the paper.

## 2. OVERVIEW OF RELATED WORK

The Young-Helmholtz scientific theory of trichromatic color vision catalyzed the emergence of the RGB color model [20]. The RGB color model is an additive tristimulus color model that describes how the visual photoreceptors and the visual cortex of the HVS facilitate the interpretation of photon waves in the following wavelength ranges: 650nm (long), 510nm (medium) and 475nm (short). As such, these long, medium and short photon waves are interpreted as red, green and blue, respectively [21]-[23]. In terms of spectral sensitivity and brightness perception, the HVS is very sensitive to medium photon waves, which are perceived as green (see Fig. 1). The RGB color model subsequently gave rise to the RGB color space for utilization on computing and visual display technologies. The interaction of the HVS with the combination of physical luminance and photon waves in nature, emitted from a display unit — which constitutes the light source — is the physical process by which visible light (photons) and the associated luminance are visually perceived as a combination of brightness, hue and saturation. On computing devices, RGB color values are typically represented as integer triplets; i.e., $2^h-1$, where $h$ denotes the bit-depth of the RGB data (bits per sample per channel). In the case of 10-bits per sample raw RGB 4:4:4 video data (i.e., 30-bit deep color), $R \in [0,1023]$, $G \in [0,1023]$ and $B \in [0,1023]$.

In terms of the relevant state-of-the-art algorithms that are present in the literature, Song et al. [13] propose a block adaptive inter-color compensation scheme for RGB 4:4:4 video coding. It reduces inter-color redundancy in RGB channels by employing a novel linear model; this method achieves coding efficiency gains by approximately 20%. Kim et al. [14] propose an inter-color redundancy technique for RGB 4:4:4 video coding in which they utilize an adaptive inter-plane weighted prediction method; this method attains coding efficiency gains of up to 21%. Zhao and Ai [15] propose a color redundancy reduction technique for RGB 4:4:4 intra coding. In this technique, the authors employ an adaptive inter-color prediction scheme whereby the B and R components are predicted from the G component; this method improves coding efficiency by around 30%.

Huang et al. [16] propose an adaptive weighted distortion scheme for utilization in the Rate-Distortion Optimization (RDO) process within HEVC. In this method, PSNR is quantified by taking the importance of the G channel — relative to the B and R channels — into account; this technique improves coding efficiency by up to 44%. Huang and Lei [17] propose a cross component technique designed for the inter-prediction process in HEVC. In this method, statistical correlations within the R, G and B color components are identified in the motion compensation prediction signal, from which a residual prediction parameter is derived; this technique achieves coding efficiency gains of up to 31%. Shang et al. [18] propose a perceptual quantization technique for HEVC. In this method, the authors develop frequency weighting matrices based on a Contrast Sensitivity Function (CSF) model, from which QMs are derived. This method is a spatial masking technique that operates at the transform coefficient level; it achieves coding efficiency gains of up to 21%.

The principle drawback of the previously proposed RGB 4:4:4 video coding techniques proposed in [13]-[17] is that they are not perceptually optimized. Though they achieve noteworthy coding efficiency improvements, the bitrate reductions could have been much greater had they taken into account the spatiotemporal perceptual redundancies inherent in the raw RGB 4:4:4 data. The QM technique proposed in [18] accounts for HVS perceptual redundancies; however, it does have certain shortcomings. The authors of this method adopt a spatial CSF model that is not designed for RGB 4:4:4 data. Furthermore, this QM technique comprises static, non-adaptive intra and inter QMs for spatial masking only. This method does not account for the characteristics of different RGB 4:4:4 video sequences, nor does it take temporal information into account.

## 3. PROPOSED SPAQ TECHNIQUE

SPAQ is a spatiotemporal perceptual quantization technique in which HVS spectral sensitivity, spatial variance and motion vector information are exploited. SPAQ is based on similar principles to our previously proposed methods known as C-BAQ [24] and FCPQ [25]. Both C-BAQ and FCPQ are spatial-only perceptual quantization techniques designed to improve upon AdaptiveQP [26]. AdaptiveQP is a perceptual quantization method that works in combination with URQ and RDOQ in HEVC to further reduce bitrates. It is designed to compute the spatial variance of Y (or G) CB sub-blocks only. Consequently, AdaptiveQP adjusts the QP of an entire $2N \times 2N$ CU without taking into account the variance of data in the sub-blocks of chroma Cb (or B) and Cr (or R) CBs, which is a significant shortcoming of the technique. With application for the perceptual coding of YCbCr data, C-BAQ [24] improves upon AdaptiveQP by taking into account the combined spatial variance of data in the sub-blocks of Y, Cb and Cr CBs; the CU-level QP is thus adjusted accordingly. FCPQ [25] expands upon C-BAQ by computing QPs at the CB level. That is, FCPQ separately adjusts the QPs for the Y CB, the Cb CB and the Cr CB based on the variances of data in the sub-blocks of all three CBs. Consequently, three CB-level QPs are signaled in the Picture Parameter Set (PPS) [27, 28]; this is also the case for SPAQ. SPAQ significantly improves upon AdaptiveQP, C-BAQ and FCPQ in the following areas: SPAQ accounts for color masking, spatial masking and temporal masking. Furthermore, AdaptiveQP, C-BAQ and FCPQ are designed to perceptually decrease the QP for smooth (low variance) regions of raw data regardless of the QStep employed during the coding process, which unnecessarily increases bitrates. Because SPAQ employs color masking and temporal masking, it is not necessary to decrease the CB-level QP in smooth regions of an RGB 4:4:4 sequence.

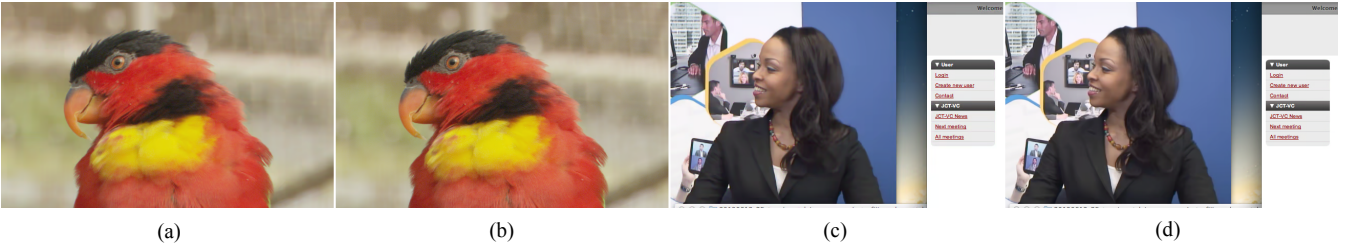

**Fig. 2**: Subfigure (a) SPAQ-coded natural content BirdsInCage video versus subfigure (b) Anchor 1-coded natural content BirdsInCage video (QP 27 test); in this test, MOS = 5 for SPAQ versus Anchor 1 and SSIM = 0.9892. Subfigure (c) SPAQ-coded screen content MissionControlClip video versus subfigure (d) Anchor 1-coded screen content MissionControlClip video (QP 27 test); in this test, MOS = 5 for SPAQ versus Anchor 2 and SSIM = 0.9891.

In terms of the SPAQ algorithm, the CB-level perceptual QPs, denoted as $Q_G$, $Q_B$, $Q_R$, are shown in (1)-(3), respectively:

$$Q_G = t_G + \left(q_G + \left[6 \times \log_2(G)\right]\right) \in \left[\frac{o}{2}, o\right] \quad (1)$$

$$Q_B = t_{B,R} + \left(q_B + \left[6 \times \log_2(B)\right]\right) \in [o, o_{max}] \quad (2)$$

$$Q_R = t_{B,R} + \left(q_R + \left[6 \times \log_2(R)\right]\right) \in [o, o_{max}] \quad (3)$$

$$o = \frac{1}{w}\sum_{k=1}^{w} p_k = 6 \quad (4)$$

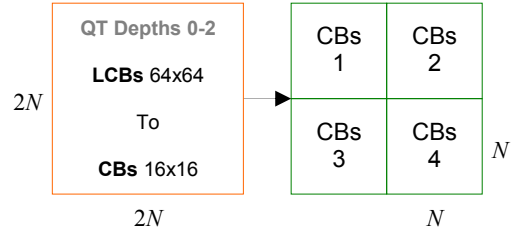

**Fig. 3**: The $2N \times 2N$ G, B and R CBs, at QuadTree (QT) depth levels 0-2, are partitioned into four $N \times N$ G, B and R CBs; $N=32$ (level 0), $N=16$ (level 1) and $N=8$ (level 2) [7]. SPAQ operates within these QT depth levels.

where $t_G$, and $t_{B,R}$ correspond to perceptual QP offsets related to temporal masking (via motion vector magnitude data in PUs) in RGB 4:4:4 video data, respectively, where $q_G$, $q_B$ and $q_R$, denote the frame-level QPs for the G, B and R channels, respectively and where $G$, $B$ and $R$ refer to variables comprising the spatial variance information in each G, B and R CB sub-block, respectively. These spatial masking-related and temporal masking-related perceptual QP offsets are employed to perceptually adjust, at the CB-level, the frame-level QP for each color channel. Variable $o$ refers to the mean CB-level perceptual QP offset in HEVC HM RExt, $o_{max}$ is the maximum allowable CB-level QP offset in HEVC HM RExt ($o_{max} = 12$ [29]), $w$ refers to the maximum number of CB-level QP offsets permitted in HEVC HM RExt ($w = 12$ [29]) and where $p_k$ refers to the $k^{th}$ CB-level QP offset. The perceptual QP offset range is lower for G data, as compared with B and R data, because humans are more sensitive to compression artifacts in G data due to the HVS spectral sensitivity phenomenon (see Fig. 1). Variables $t_G$ and $t_{B,R}$ are computed in (5)-(8), respectively. $G$, $B$ and $R$, and the associated variables, are quantified in (9)-(11), respectively:

$$t_G = \begin{cases} \frac{o}{2} & \text{if } M(n,i) > V, \\ 0 & \text{else.} \end{cases} \quad (5)$$

$$t_{B,R} = \begin{cases} o & \text{if } M(n,i) > V, \\ 0 & \text{else.} \end{cases} \quad (6)$$

$$V = \frac{1}{P}\sum_{i=1}^{P} M(n,i) \quad (7)$$

$$M(n,i) = \sqrt{MV_x^2 + MV_y^2} \quad (8)$$

where $M(n,i)$ corresponds to the magnitude of motion vector $MV$ in a PU and where $V$ is the arithmetic mean motion vector magnitude in a PU within an entire frame. More specifically, $M(n,i)$ denotes the magnitude of a motion vector in a PU within the $i^{th}$ CU of the $n^{th}$ frame and where $P$ corresponds to the total number of PUs in the $n^{th}$ frame. Subscripts $x$ and $y$ correspond to the coordinates $(x,y)$ of motion vector $MV$. $M(n,i)$ constitutes an adaptive threshold value related to temporal masking, whereby magnitude $M(n,i)$ of motion vector $MV$ must exceed $V$ in order for a region to be considered as high motion. Note that, because the G, B and R CBs are of equal size (see Fig. 3), the motion vectors in each G, B and R Prediction Block (PB) do not differ in magnitude (i.e., no scaling is required), which is the reason why SPAQ operates at the PU level in relation to measuring motion vector magnitudes. In terms of perceptual quantization adjustments as a result of temporal masking, recall that the HVS is much less sensitive to compression artifacts in high motion regions of video data [30]. We now define $G$, and the associated variables, in (9)-(11), respectively:

$$G = \frac{s \cdot g + m_G}{g + s \cdot m_G} \quad (9)$$

$$g = 1 + \min(\sigma^2_{G,d}), \quad \text{where } d = 1,...,4 \quad (10)$$

$$m_G = \frac{1}{C_G}\sum_{j=1}^{C_G} g_j \quad (11)$$

where variable $s$ refers to a scaling factor for normalizing the spatial activity of a G CB ($s = 2$ by default in HEVC [29]), where $g$ corresponds to the non-normalized variance of a G CB, where $m_G$ denotes the mean variance of all $2N \times 2N$ G CBs belonging to the current picture of a frame, where $\sigma^2_{G,d}$ refers to the variance of samples in an $N \times N$ CB sub-block $d$ within a G CB and where $C_G$ denotes the number of $2N \times 2N$ G CBs in the current picture. Note that the equations for quantifying the variances of $B$ and $R$, from Eqs. (2) and (3), respectively, are essentially identical to those shown in Eqs. (9)-(11), which calculate $G$ from Eq. (1). Only the symbols that express the numerical data for $B$ and $R$ differ.

**Table 1**: Bitrate reductions achieved by SPAQ compared with anchors over QPs 22, 27, 32 and 37. The decreased objective reconstruction quality incurred by SPAQ is quantified by G, B and R channel PSNR reductions (%). Visual quality is quantified by the subjective MOS and perceptual SSIM scores.

| Bitrate Reductions (%), G, B and R PSNR Reductions (%), Global GBR SSIM Scores and MOS Scores — All Averaged Over QPs 22, 27, 32 and 37 | | | | | | | | | | | | | |
|---|---|---|---|---|---|---|---|---|---|---|---|---|---|
| **Natural Content** | **SPAQ versus HM 16.17 (Anchor 1)** | | | | | | **Screen Content** | **SPAQ versus HM 16.17 + SCM 8.7 (Anchor 2)** | | | | | |
| 30-Bit Sequence | **Bitrate** | G PSNR | B PSNR | R PSNR | SSIM | MOS | 24-Bit Sequence | **Bitrate** | G PSNR | B PSNR | R PSNR | SSIM | MOS |
| BirdsInCage | −71.7% | −0.8% | −4.0% | −5.6% | 0.9895 | 4.75 | BasketballScreen | −39.7% | −4.6% | −17.8% | −18.2% | 0.9696 | 4.50 |
| CrowdRun | −44.2% | −1.5% | −10.6% | −10.8% | 0.9450 | 4.00 | MissionControlClip | −31.7% | −4.6% | −18.7% | −19.6% | 0.9779 | 4.50 |
| DuckAndLegs | −64.4% | −1.9% | −9.8% | −9.0% | 0.9061 | 4.50 | CADWaveform | −27.6% | −6.1% | −22.4% | −22.2% | 0.9800 | 5.00 |
| Kimono | −58.8% | −1.2% | −4.9% | −5.9% | 0.9360 | 4.25 | Desktop | −30.3% | −7.2% | −24.4% | −24.6% | 0.9789 | 4.75 |
| OldTownCross | −71.7% | −1.0% | −6.7% | −6.0% | 0.8755 | 4.75 | FlyingGraphics | −34.8% | −2.5% | −16.3% | −15.7% | 0.9829 | 5.00 |
| ParkScene | −60.9% | −1.6% | −6.5% | −8.4% | 0.9123 | 4.25 | PPT_DOC_XLS | −31.6% | −5.3% | −21.4% | −22.4% | 0.9697 | 5.00 |
| Seeking | −54.7% | −1.2% | −8.1% | −9.0% | 0.9361 | 3.75 | SocialNetworkMap | −39.3% | −1.8% | −14.6% | −15.1% | 0.9617 | 4.75 |
| Traffic | −52.4% | −2.4% | −9.2% | −11.0% | 0.9386 | 4.25 | VenueVu | −37.3% | −0.9% | −7.6% | −7.9% | 0.9778 | 4.50 |

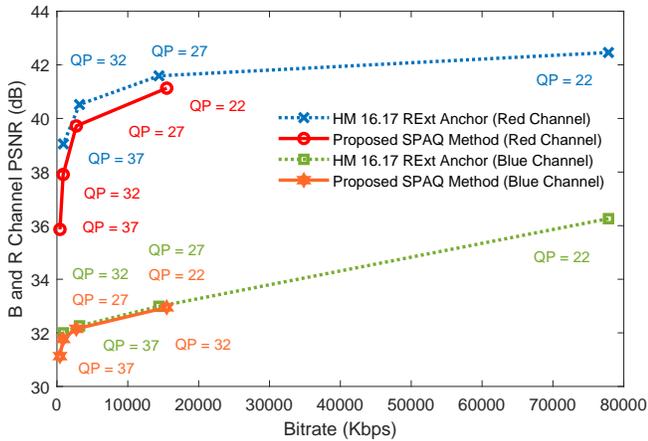

**Fig. 4**: A plot showing the bitrate reductions attained by SPAQ (B and R color channels) over four QP data points compared with HM 16.17 RExt on the BirdsInCage RGB 4:4:4 30-bit video sequence (RA configuration).

## 4. EVALUATION, RESULTS AND DISCUSSION

We implement the proposed SPAQ technique into the JCT-VC HEVC HM 16.17 RExt + SCM 8.7 codec [29]. SPAQ is compared with two anchors, which are as follows: HEVC HM 16.17 RExt (for the coding of RGB 4:4:4 natural content) and also HEVC HM 16.17 RExt + SCM 8.7 (for the coding of RGB 4:4:4 screen content); AdaptiveQP is enabled in anchors. The evaluations are conducted on 16 official JCT-VC RGB 4:4:4 video sequences, which are listed in Table 1. The natural content is 10-bits per sample and the screen content is 8-bits per sample; these sequences have a resolution of HD 1080p. In the evaluation, we measure the bitrate reductions SPAQ versus anchors over four QP data points (QPs 22, 27, 32 and 37) using the Random Access (RA) configuration as per the JCT-VC common test conditions specified in [31, 32]. In order to ascertain the perceptual compression efficacy of SPAQ versus anchors, the subjective evaluations — quantified using MOS as per ITU-R Rec. P.910 [33] — and SSIM [34] are the most important perceptual metrics. MOS = 5 equates to imperceptible distortion, MOS = 4 equates to near-imperceptible distortion and MOS = 3 equates to perceptible distortion, but not overly distracting. MOS < 3 usually corresponds to poor subjective quality. As regards the SSIM metric, a score of SSIM > 0.95 is often considered to correlate with MOS = 5. SSIM = 1 is the maximum value (i.e., mathematically lossless reconstruction). The subjective tests are conducted on a HD 50 inch display at a viewing distance of 0.75m ≈ 29.5 inch. In line with ITU-R Rec. P.910 [33], four participants engaged in the subjective evaluations; i.e., by analyzing sequences coded using QPs 22, 27, 32 and 37.

In comparison with anchors (see Table 1), SPAQ attains vast bitrate reductions in all tests (i.e., initial QPs 22, 27, 32 and 37). The most noteworthy bitrate reductions achieved by SPAQ are as follows: 80.1% (QP 22 test) and 81% (QP 27 test) on the BirdsInCage sequence (see Table 1, Fig. 2 and Fig. 4). In terms of perceptual quality, the subjective evaluation participants chose MOS = 5 in both the QP 22 and QP 27 tests. This means that the SPAQ-coded BirdsInCage sequence proved to be visually identical to the BirdsInCage sequence coded by anchors. The SSIM scores in these tests correlated with the subjective score of MOS = 5. That is, SSIM = 0.9899 (QP 22 test) and SSIM = 0.9892 (QP 27 test). SPAQ proved to be ineffective in the majority of the QP 37 tests. This is because QP = 37 is a very high initial QP, thus leaving little room to further reduce psychovisual redundancies by virtue of perceptual QP offsets. Regarding overall MOS and SSIM scores for the 16 sequences shown in Table 1, an MOS = 5 was scored by each participant in 100% of QP 22 tests, 94% of QP 27 tests, 44% of QP 32 tests and 19% of QP 37 tests. An SSIM > 0.95 was recorded in the vast majority of QP 22 and QP 27 tests.

As shown in Table 1, the mathematical reconstruction quality — as quantified by G, B and R PSNR (dB) percentage values — is considerably inferior for SPAQ. This is due to the higher levels of quantization applied to B and R data, in particular, in accordance with color masking, spatial masking and temporal masking based on a HVS model. However, these substantial PSNR dB reductions have little correlation with subjective MOS and SSIM scores, other than when the initial QP is very high (e.g., in the QP = 37 tests). In the evaluations for SPAQ versus anchors, we have proved that high levels of perceptual quantization can be applied to G, B and R components without incurring visually conspicuous compression artifacts; to reiterate, this proved to be the case in the QP 22 and QP 27 tests (and certain QP 32 tests). Moreover, the evidence provided in these experiments reaffirms the fact that the PSNR metric has a poor correlation with human visual perception.

## 5. CONCLUSION

In this paper, we propose a spatiotemporal perceptual quantization technique, named SPAQ, for application with RGB 4:4:4 video data. We exploit HVS spectral sensitivity, color masking, spatial masking and temporal masking to discard perceptually irrelevant samples from each sequence. Compared with anchors, SPAQ achieves vast bitrate reductions, of up to 81%, while achieving perceptually lossless compression (in the QP 22 and QP 27 tests). In a side test, further subjective evaluations were undertaken versus the raw RGB 4:4:4 sequences; MOS = 5 was chosen by the participants for SPAQ on all sequences (QP = 22 test only). SPAQ thus achieves perceptually lossless coding, versus the raw data, at QP = 22. Finally, SPAQ attains moderate encoding time and decoding time improvements compared with anchors.